\documentclass{MYws-ijmpa}
\usepackage[super,compress]{cite}
\usepackage{graphicx}
\usepackage{xcolor}
\usepackage{hyperref}
\usepackage{url}
\begin{document}
\markboth{German Valencia}{Pursue of CP violation in hyperon decay}

\catchline{}{}{}{}{}
%
%%%%%%%%%%%%%%%%%%%%%%%%%%%%%%%%%%%%%%%%%%%%%%%%%%%%%%%%%%%%%%%%%%%%

\title{Pursue of CP violation in hyperon decay}

\author{German Valencia}

\address{School of Physics and Astronomy, Monash University,\\
  Wellington Road, Clayton, VIC-3800, Australia\\
German.Valencia@monash.edu}

\maketitle

%\begin{history}
%\received{Day Month Year}
%\revised{Day Month Year}
%\end{history}

\begin{abstract}
I review the status of CP violation in hyperon decay in light of recent progress by BESIII and the anticipated improvements at the super tau-charm facility. I emphasize the complementarity between kaons and hyperons for studying CP violation in $|\Delta S|=1,2$ processes.
\end{abstract}

\ccode{PACS numbers:11.30.Er 14.20.Jn 13.30.Eg}

%\tableofcontents

\section{CP violation in decays of strange hadrons}	

The study of CP violation in kaon decay has a long and rich history that started with the first observation of $K_L\to \pi\pi$ in 1964 \cite{Christenson:1964fg}, the attribution of its origin to kaon mixing and ultimately to the phase in the CKM matrix within the SM \cite{Cabibbo:1963yz,Kobayashi:1973fv}. The first evidence for direct CP violation, came with the measurement of $\epsilon^\prime/\epsilon$ by NA31 in 1988 \cite{NA31:1988eyf} and was later confirmed by NA48 and KTeV in the early 2000s \cite{NA31:1993tha,KTeV:1999kad,NA48:2002tmj}.

From the theoretical perspective, calculations of the parameters $\epsilon$ and $\epsilon^\prime$ have been carried out for over forty years. These calculations are notoriously difficult and even now, large uncertainties remain \cite{Brod:2019rzc,Cirigliano:2019cpi}. 

CP violation in hyperon decay is probably harder to estimate than its counterpart in the $K\to \pi\pi$ system, and unfortunately, it has received much less attention. The recent experimental progress by BESIII should inspire new theoretical efforts, and an eventual observation of CP violation in hyperon non-leptonic decay would help us close the window on new physics that may affect strange hadron decay.

\begin{figure}[!htb]
\centerline{\includegraphics[width=1.0\linewidth]{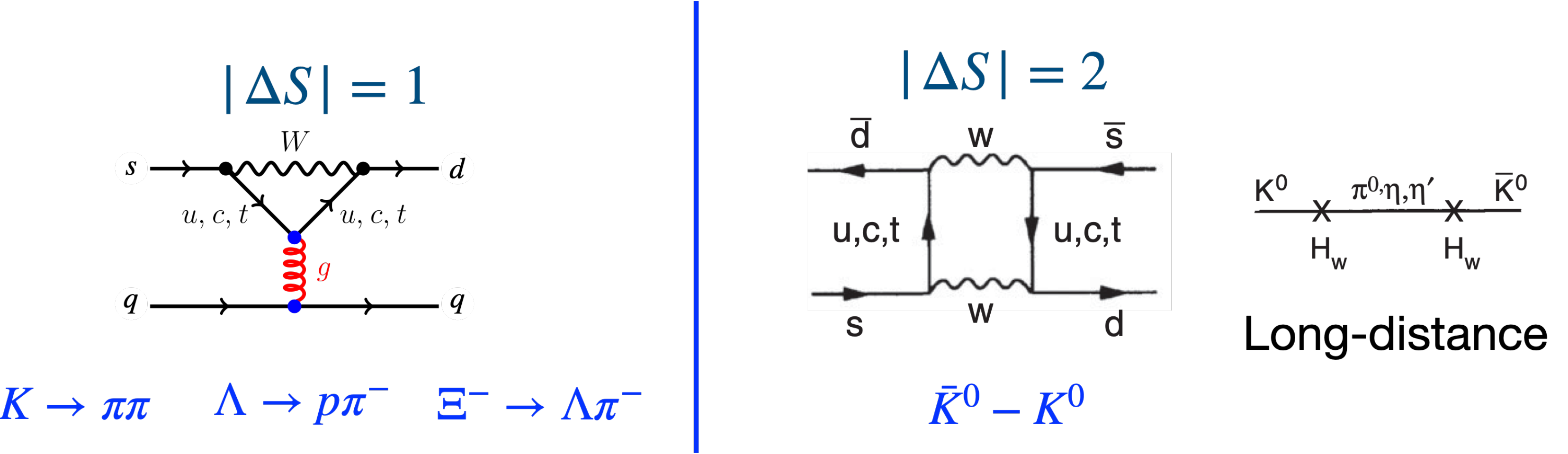}}
\caption{Diagrams illustrating the sources of CP violation in the SM for $|\Delta S|=1$ processes (left panel) and $|\Delta S|=2$ processes including both short and long-distance contributions  (right panel). \label{f1}}
\end{figure}

Within the SM, the CP violating phase present in the CKM matrix enters both $|\Delta S|=1$ processes such as $K\to \pi\pi$, $\Lambda\to p\pi^-$, and $\Xi^-\to \Lambda \pi^-$  as depicted in the left panel of Fig.~\ref{f1} and $|\Delta S|=2$ processes such as kaon mixing depicted on the right panel of the same figure.

All these processes then probe the same quantity, $\sim{\rm Im}(V_{td}V^*_{ts})=A^2\lambda^5\eta$ in the Wolfenstein parametrization of the CKM matrix \cite{Wolfenstein:1983yz}.

The measurements of kaon observables have established CP violation, but the uncertainty in their calculation still allows for relatively large contributions beyond the SM. With general new physics, the kaon and hyperon modes are complementary and observing CP violation in hyperon decay would add valuable information to our overall picture.

\subsection{CP violation in $K\to \pi\pi$}

The parameters $\epsilon$ and $\epsilon^\prime$ characterizing indirect and direct CP violation respectively, are extracted from measurements of $K\to \pi\pi$ modes according to
\begin{align}
    \eta^{+-}=&\frac{A(K_L\to\pi^+\pi^-)}{A(K_S\to\pi^+\pi^-)}=\epsilon+\epsilon^\prime,\\ \eta^{00}=&\frac{A(K_L\to\pi^0\pi^0)}{A(K_S\to\pi^0\pi^0)}=\epsilon-2\epsilon^\prime.
\end{align}
The current experimental averages quoted by the PDG \cite{ParticleDataGroup:2022pth} along with recent theoretical calculations are compared in Table~\ref{t0}. The last column of that table shows the difference between experiment and theory, the current window for new physics.

\begin{table}[!htb]
\tbl{Current experimental and theoretical status of $\epsilon$ and $\epsilon^\prime$.}
{\begin{tabular}{@{}cccc@{}}\hline
 & Experiment \cite{ParticleDataGroup:2022pth} & Theory \cite{Brod:2019rzc,Cirigliano:2019cpi} & New physics window \\ \hline
 $|\epsilon|$ & $(2.228\pm0.011)\times 10^{-3}$ & $(2.16\pm0.18 )\times 10^{-3}$  & $\left|\epsilon\right|_{BSM}=(0.7\pm 1.8)\times 10^{-4}$\\
 ${\rm Re}\left(\frac{\epsilon^\prime}{\epsilon}\right)$ & $(1.66\pm 0.23)\times 10^{-3}$  & $(1.3^{+0.6}_{-0.7} )\times 10^{-3}$  & $\left|\frac{\epsilon^\prime}{\epsilon}\right|_{BSM}=(0.4^{+0.7}_{-0.6})\times 10^{-3}$\\
 \hline
\end{tabular}\label{t0}}
\end{table}  

\subsection{CP violating observables in hyperon non-leptonic decay}

\begin{figure}[!htb]
\centerline{\includegraphics[width=1.0\linewidth]{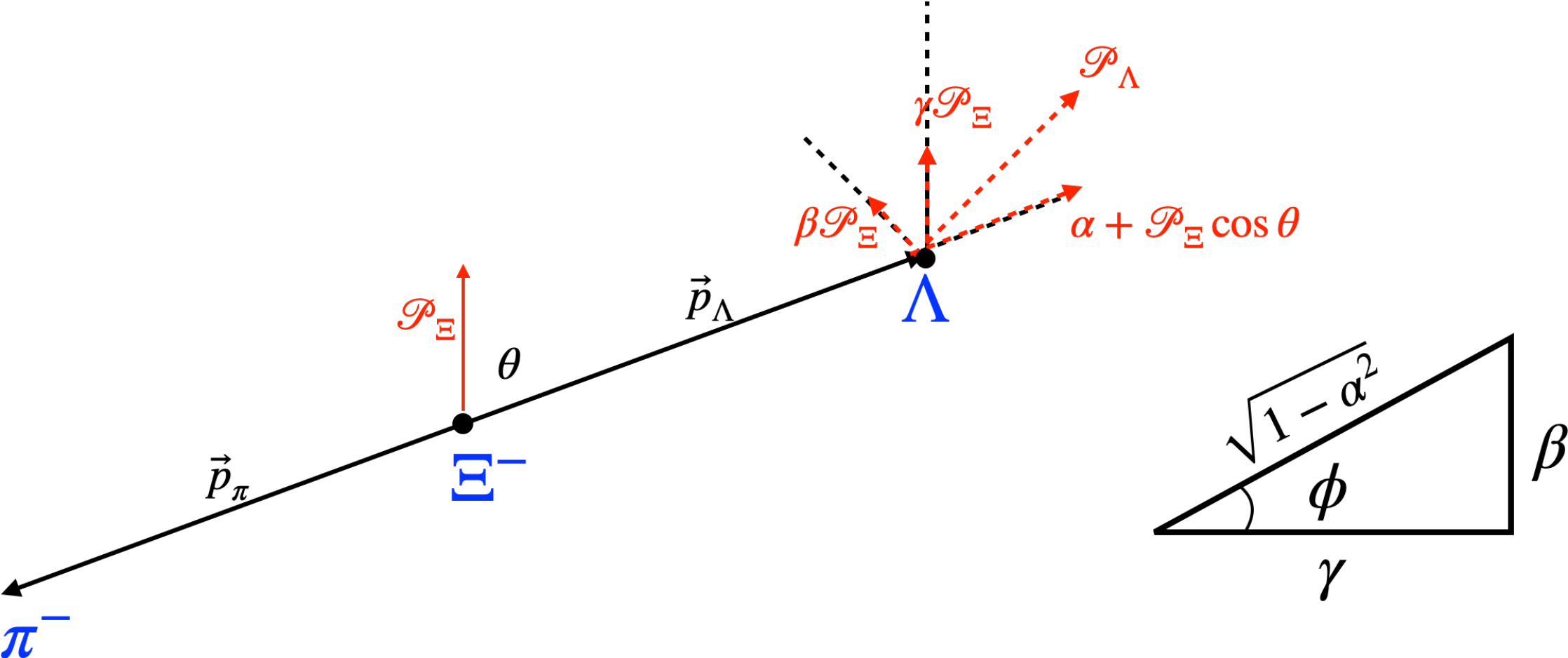}}
\caption{Different components of ${\cal P}_f$, the polarization of the daughter hyperon, and their relation to the parameters $\alpha$, $\beta$ and $\gamma$ for the case of $\Xi^-\to \Lambda \pi^-$. On the right, we illustrate the relation between parameters $\alpha,\beta,\gamma$ and $\phi$.\label{f2}}
\end{figure}

The angular distribution for the decay of a polarized hyperon with polarization ${\cal P}_i$ into a daughter baryon  and a pion, ${\cal B}_i({\cal P}_i)\to {\cal B}_f(\vec{p}_f,{\cal P}_f)\pi $,  is given in terms of a parameter $\alpha$ by,
\begin{align}
    \frac{d\Gamma_{{\cal B}_i\to {\cal B}_f\pi}}{d\Omega_f}  &= \frac{\Gamma_{{\cal B}_i\to {\cal B}_f\pi}}{4\pi}\left(1+\alpha\, {\bf {\cal P}}\!_i\cdot\hat{\bf p}_{\!f}\right).
\end{align}
The final state baryon emerges with polarization  ${\cal P}_f$ which is given in terms of $\alpha$ and additional parameter $\beta$ by
\begin{align}
    \bf{{\cal P}}_f&=\frac{\left(\alpha+\bf{{\cal P}}_i\cdot\bf{\hat p}_f\right)\bf{\hat p}_f+\beta\bf{{\cal P}}_i\times\bf{\hat p}_f+\gamma\bf{\hat p}_f\times(\bf{{\cal P}}_i\times\bf{\hat p}_f)}{1+\alpha\bf{{\cal P}}_i\cdot\bf{\hat p}_f} 
    \label{eq:pol}
\end{align}
The last term is written in terms of $\gamma$, where $\alpha^2+\beta^2+\gamma^2=1$. It is also customary to use a fourth parameter $\phi$, and the different components of ${\cal P}_f$  and their relation to these parameters are illustrated in Fig.~\ref{f2} for the case of $\Xi^-\to \Lambda \pi^-$.

For theoretical predictions, it is convenient to decompose the decay amplitudes into isospin and parity components \cite{Brown:1983wd,Donoghue:1985ww}. For example, the amplitude for the decay $\Lambda \to p\pi^-$ has $S$ and $P$ waves, and the isospin of the final state can be $1/2$ or $3/2$. The transition matrix element for this decay  can then be written as,
\begin{align}
    {\cal M} =& \chi_f^\dagger\left(S+{\mathbf \sigma\cdot\hat{\mathbf p}_f ~P}\right)\chi_i\\
    S=& S_1e^{i(\delta_1^S+\xi_1^S)}+S_3e^{i(\delta_3^S+\xi_3^S)}\\
    P= & P_1e^{i(\delta_1^P+\xi_1^P)}+P_3e^{i(\delta_3^P+\xi_3^P)}.
\end{align}
In these expressions, the strong rescattering phases are explicitly written as $\delta$ and the weak, CP-violating phases as $\xi$.  The notation $\chi_{i,f}$ refers to the two-component spinor for the initial and final state fermion respectively, and the indices $(1,3)$ denote twice the isospin of the final $p\pi^-$ state. The $S$ and $P$ amplitudes exhibit a hierarchy $S_3<<S_1$ and $P_3<<P_1$, the $\Delta I =1/2$ rule, and can be extracted from experiment. The available information for two of the hyperon decay modes is shown in Table~\ref{ta1}, with the $S$ and $P$ waves extracted from a fit to branching ratios and values of $\alpha$ taken from the PDG \cite{ParticleDataGroup:2022pth}. The same fit reveals the relative smallness of $\Delta I =3/2$ transitions as seen in the last two columns. 

\begin{table}[!htb]
\tbl{Values of S and P waves for two examples of hyperon non-leptonic decay extracted from a recent fit.}
{\begin{tabular}{|l|c|c|c|c|}\hline
mode & $S$ & $P$ & $S_3/S_1$& $P_3/P_1$ \\ \hline
$\Lambda\to p\pi^-$ & $1.382\pm0.008$ & $0.624\pm0.005$ & $0.03\pm0.01$  & $-0.04\pm0.02$ \\
$\Xi^-\to \Lambda \pi^-$ &$-1.994\pm0.009$ & $0.392\pm0.004$ & $0.04\pm0.01$ & $0.01\pm0.02$ \\ \hline
\end{tabular}\label{ta1}}
\end{table}  

The strong phases in $\Lambda$ decay, are the pion-nucleon scattering phases and have been extracted from \cite{Hoferichter:2015hva}, $(\delta_1^S-\delta_1^P)=  7.31^\circ\pm0.12^\circ$. The strong phases in $\Xi^-$ decay have been measured by HyperCP \cite{HyperCP:2004not} $(\delta_2^P-\delta_2^S)= 4.6^\circ\pm1.8^\circ$\footnote{BESIII also has a result with larger uncertainty \cite{BESIII:2021ypr}.} and have also been calculated in $\chi$PT, with a recent result being $\delta^P-\delta^S=(8.78^{+0.19}_{-0.22})^\circ$ \cite{Huang:2017bmx}.

\section{Tests of CP violation}

CP invariance can be tested by comparing a hyperon decay such as $\Lambda \to p \pi^-$ to its corresponding antihyperon decay $\bar\Lambda \to \bar p \pi^+$. If CP invariance holds then the decay rates are equal,  $\Gamma= \bar\Gamma$, and the decay parameters satisfy $\alpha=-\bar\alpha,~~\beta=-\bar\beta$. This offers the following observables as tests of CP invariance, 
\begin{align}
    \Delta_{CP}  =& \frac{\Gamma-\overline\Gamma}{\Gamma+\overline\Gamma}\\
    A_{CP}^{}  =& \frac{\alpha+\overline\alpha}{\alpha-\overline\alpha} \\
    B_{CP}^{}  =& \frac{\beta+\overline\beta}{\alpha-\overline\alpha}
    \label{CPobs}
\end{align}
Due to the smallness of $\Delta I=3/2$ amplitudes, the strong scattering phases, and the weak CP violating phases, these observables can be expanded to leading order in all three quantities to obtain relatively simple expressions \cite{Donoghue:1985ww,Donoghue:1986hh}. For example, for $\Lambda\to p\pi^-$ one finds
\begin{align}
     \Delta_{CP}\simeq & \sqrt{2}\underbrace{\frac{S_3}{S_1}}_{\Delta I =1/2{\rm ~rule}}\underbrace{\sin(\delta_3^S-\delta_1^S)}_{\rm strong~phases}\underbrace{\sin(\xi_3^S-\xi_1^S)}_{\rm weak~phases} \\ \\
     A_{CP}^{} \simeq & -\tan(\delta^P-\delta^S)\,\tan(\xi^P-\xi^S) \\ 
   B_{CP}^{}  \simeq &\tan(\xi^P-\xi^S),
\end{align}
and therefore $B_{CP}>A_{CP}>\Delta_{CP}$. $B_{CP}$ probes for an asymmetry in the parameter $\beta$ from Eq.~(\ref{eq:pol})  which does not require a strong phase as it originates from a $T$-odd triple product correlation in the angular distribution of the decaying hyperon proportional to $\vec{p}_\pi.({\cal P}_f\times{\cal P}_i)$. The normalization of $B_{CP}$ is chosen to be $\alpha-\bar \alpha$ and not $\beta-\bar\beta$ to better reflect the difficulty of its measurement \cite{Donoghue:1985ww,Donoghue:1986nn}.

At BES and BESIII the hyperons are produced in pairs (the same will be true at SCTF) in $e^+e^-$ collisions in reactions with sequential decays such as
\begin{align}
    e^+e^-\to J/\psi\to \Xi^-\bar\Xi^+ \to \Lambda\bar\Lambda \pi^+\pi^-\to p\bar p \pi^+\pi^-\pi^+\pi^-
    \label{eqsec}
\end{align}
where the weak decays of the hyperons analyze their polarization. The combined angular distribution for a process like this one allows BESIII to simultaneously extract the parameters $\alpha_\Xi,\bar\alpha_\Xi,\alpha_\Lambda,\bar \alpha_\Lambda,\phi_\Xi,\bar \phi_\Xi$ \cite{Perotti:2018wxm,BESIII:2021ypr}. 

With samples of $1.3\times 10^9,~1\times 10^{10}$, and $3.4\times 10^{12}$ $J/\psi$ collected at BESIII and projected at SCTF respectively, statistical sensitivities to the various observables have been estimated to range from $10^{-2}$ to $2.3\times 10^{-4}$ \cite{Salone:2022lpt}. Some of these sensitivities are summarised in Table~\ref{ta3}.

\begin{table}[!htb]
\tbl{Selected statistical sensitivities to $A_{CP}^{\Xi}$ taken from \cite{Salone:2022lpt}}
{\begin{tabular}{@{}|c|c|c|@{}}\hline
& Number of $J/\psi$ & sensitivity to $A_{CP}^{\Xi}$ \\ \hline
BESIII  & $1.3\times 10^9$ & $1.3\times 10^{-2}$ \\
BESIII  & $1\times 10^{10}$ & $4.8\times 10^{-3}$ \\
tau-charm factory & $3.4\times 10^{12}$ & $2.6\times 10^{-4}$ \\ \hline
\end{tabular}\label{ta3}}
\end{table}      

The existing experimental results for $A_{CP}^\Lambda$ are summarized in Table~\ref{ta4}, whereas those for other hyperon observables are summarized in Table-\ref{ta5}.

\begin{table}[!htb]
\tbl{Existing limits on $A^\Lambda_{CP}$.}
{\begin{tabular}{@{}|c|c|c|c|@{}}\hline
 &  & process & Experiment \\ \hline
 $A_{CP}^\Lambda$ & $-0.004\pm 0.012\pm 0.009$ & $J/\Psi\to \Xi\bar \Xi \to\Lambda \bar \Lambda \pi\pi$ & BESIII (2022)\cite{BESIII:2021ypr}\\
  $A_{CP}^\Lambda$ & $-0.0025\pm 0.0046\pm 0.0012$ & $J/\Psi\to\Lambda \bar \Lambda $ & BESIII (2022)\cite{BESIII:2022qax} \\
  $A_{CP}^\Lambda$ & $-0.081\pm 0.055\pm 0.059$ & $J/\Psi\to\Lambda \bar \Lambda $ & BES (2010)\cite{BES:2009zvb} \\ 
  $A_{CP}^\Lambda$ & $0.013\pm 0.022$ & $p\bar p\to\Lambda \bar \Lambda $ & LEAR (1996)\cite{Barnes:1996si} \\
   $A_{CP}^\Lambda$ & $0.01\pm 0.10$ & $J/\Psi\to\Lambda \bar \Lambda $ &DM2 (1988)\cite{DM2:1988ppi}\\ 
  $A_{CP}^\Lambda$ & $-0.002\pm 0.004$ &  &PDG average\cite{ParticleDataGroup:2022pth}\\ \hline
\end{tabular}\label{ta4}}
\end{table}      

\begin{table}[!htb]
\tbl{Other existing limits on CP-violating observables in hyperon decay.}
{\begin{tabular}{@{}|c|c|c|c|@{}}\hline
 &  & process & Experiment \\ \hline
   ${\color{red} B_{CP}^\Xi\approx}~(\xi_P-\xi_S)^\Xi$ & $(1.2\pm3.4\pm0.8)\times 10^{-2}$~rad & $J/\Psi\to \Xi\bar \Xi \to\Lambda \bar \Lambda \pi\pi$ & BESIII (2022)\cite{BESIII:2021ypr} \\
 $A_{CP}^\Xi$ & $(-9\pm 8^{+7}_{-2})\times 10^{-3}$ & $J/\Psi\to \Xi\bar \Xi \to\Lambda \bar \Lambda \pi\pi$ & BESIII (2024)\cite{BESIII:2023jhj}\\
 $A_{CP}^\Xi$ & $(-1.5\pm5.1\pm1.0)\times 10^{-2}$ & $\Psi(3686)\to \Xi\bar \Xi \to\Lambda \bar \Lambda \pi\pi$ & BESIII (2022)\cite{BESIII:2022lsz}\\
  $A_{CP}^{\Xi^0}$ & $(-5.4\pm6.5\pm3.1)\times 10^{-3}$ & $J/\Psi\to \Xi\bar \Xi \to\Lambda \bar \Lambda \pi\pi$ & BESIII (2023)\cite{BESIII:2023drj}\\
  $A_{CP}^\Lambda+A_{CP}^\Xi$ & $(0.0\pm5.1\pm4.4)\times 10^{-4}$ & $\Xi\to\Lambda\pi\to p\pi\pi $ & HyperCP (2004)\cite{HyperCP:2004zvh}\\
   $A_{CP}^{\Omega\to\Lambda K}$ & $-0.016\pm0.092\pm0.089$ & $\bar\Omega^+\to\bar \Lambda K^+\to \bar p\pi^+ K^+$ & HyperCP (2006)\cite{HyperCP:2006ktj} \\
    $A_{CP}^{\Sigma^+}$ & $0.004\pm0.037\pm0.010$ & $J/\Psi/\Psi(2S)\to \Sigma^+\bar\Sigma^-\to p \bar p \pi^0\pi^0$ & BESIII (2020)\cite{BESIII:2020fqg}\\
     $A_{CP}^{\Sigma^+}$ & $-0.080\pm0.052\pm0.028$ & $J/\Psi/\Psi(2S)\to \Sigma^+\bar\Sigma^-\to n \bar n \pi^+\pi^-$ & BESIII (2023)\cite{BESIII:2023sgt}\\ \hline
\end{tabular}\label{ta5}}
\end{table}      

\subsection{Triple product correlations}

\begin{figure}[!htb]
\centerline{\includegraphics[width=1.0\linewidth]{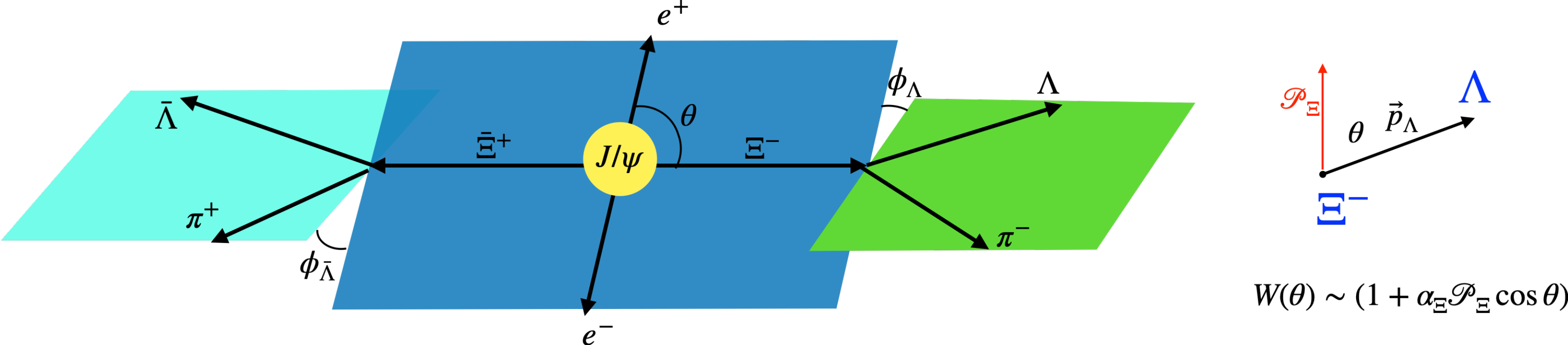}}
\caption{Polarization and momentum vectors for $e^+e^-\to J/\psi\to \Xi^-\bar\Xi^+ \to \Lambda\bar\Lambda \pi^+\pi^-\to p\bar p \pi^+\pi^-\pi^+\pi^-$ sketched on the left. The correlation between ${\cal P}_\Xi$ and $\vec{p}_\Lambda$ is determined by the  $\alpha$ parameter.  \label{ftrip}}
\end{figure}

In sequential decays for pair-produced hyperons such as the one in Eq.~(\ref{eqsec}), it is possible to construct triple product correlations that isolate the CP odd observables in the angular distribution \cite{Donoghue:1986nn}.
The connection between these triple products and the observables of Eq.~(\ref{CPobs}) is sketched in Fig.~\ref{ftrip}.\footnote{ This figure is adapted from those in \cite{Schonning:2023mge}.}

Following Fig.~\ref{ftrip} for the example of Eq.~(\ref{eqsec}), we first note that in the $e^+e^-$ rest frame, $\vec{p}_{\Xi^-}=-\vec{p}_{\bar \Xi^+}$ and that the polarization ${\cal P}_\Xi$ is perpendicular to the production plane (so in the direction of  $\vec{p}_{e^-}\times \vec{p}_{\Xi^-}$) due to parity conservation in the production process. As the right panel of Fig.~\ref{ftrip} reminds us, the parameter $\alpha_\Xi$ measures the correlation between the $\Xi^-$ polarization and the $\Lambda$ momentum. We then see that
\begin{align}
    {\cal O}\equiv \vec{p}_e\times\vec{p}_\Xi\cdot(\vec p_\Lambda+\vec p_{\bar \Lambda})\propto (\alpha_\Xi+\alpha_{\bar \Xi})\sim A_{CP}^\Xi
\end{align}
which can be measured with the counting asymmetry
\begin{align}
    \frac{N_{\rm ev}({\cal O}>0)-N_{\rm ev}({\cal O}<0)}{N_{\rm ev}({\cal O}>0)+N_{\rm ev}({\cal O}<0)}.
\end{align}
These counting asymmetries can also be useful for $p\bar  p$ \cite{Barnes:1996si,Barnes:1987vc,PANDA:2020zwv} production of hyperon pairs.

\section{Estimate of asymmetries}

Within the SM the phases that give rise to CP violation in $|\Delta S|=1$ transitions arise from one-loop penguin diagrams leading to complex Wilson coefficients in the effective weak Hamiltonian at the hadronic scale that multiply four-quark operators $Q_i$ as
\begin{align}
    &{\cal L}^{\Delta S=1}_{\rm eff}=-\frac{G_F}{\sqrt{2}}V_{ud}V^*_{us}\sum_{i=1}^{10} \left(z_i-\frac{V_{td}V^*_{ts}}{V_{ud}V^*_{us}}y_i\right)Q_i\\
    &Q_2=(\bar s u)_{V-A}(\bar u d)_{V-A},~Q_6=(\bar s_id_j)_{V-A}\sum_q(\bar q_j q_i)_{V+A},\\
    &Q_8=\frac{3}{2}(\bar s_id_j)_{V-A}\sum_qe_q(\bar q_j q_i)_{V+A},~~\cdots
    \label{quarkeffH}
\end{align}
Where we have only listed one tree-level operator ($Q_2$) and the main gluon  ($Q_6$) and electroweak penguin operators  ($Q_8$). 
In principle one then needs to calculate the matrix elements of the form $\left<p\pi^-\left|Q_i\right|\Lambda\right>$ to get all the amplitudes. As is well known, this does not give the correct amplitudes mostly because long-distance effects get in the way.

We can start instead with a low-energy hadron-level effective interaction. At leading order in $\chi$PT one has \cite{Bijnens:1985kj,Jenkins:1991ne},
\begin{align}
    {\cal L}_{\Delta S=1}^{SM} & \,\supset\, {\rm Tr}\left( {\color{red}h_D^{}}\, \overline B\big\{ \xi^\dagger\hat\kappa\xi,B \big\} + {\color{red}h_F^{}}\, \overline B\left[\xi^\dagger\hat\kappa\xi,B\right] \right) + {\color{red}h_C^{}}\, \left(\overline T_{kln}\right)^\eta\, \left(\xi^\dagger\hat\kappa\xi\right)_{no}\, (T_{klo})_\eta^{} 
\end{align}
where $B$ is the baryon octet, $T$ the baryon decuplet, and $\xi=e^{i\pi/f}$ contains the pion fields \footnote{For a recent extraction of these low energy constants see \cite{He:2023cqg}.}

The low energy constants ${\color{red}h_D,~h_F,~h_C}$ are extracted from fits to hyperon weak non-leptonic decay and to $\Omega$ strong decay, and suffer from the well-known S/P problem. When the S-waves are used for the fit, the values of $h_D,~h_F$ do not reproduce the observed P-waves, and vice versa. This situation is partially understood because one-loop corrections to the leading order result are large, and the discrepancy with experiment is consistent with the size of these corrections. In practice, however, going beyond leading order in $\chi$PT does not help the calculations because there are too many unknown low energy constants.

Given this situation, and the absence of lattice calculations, we adopt the following prescription to estimate the phases:
\begin{itemize}
    \item The imaginary (CP-violating) part of the amplitude originates in short-distance physics and enters the amplitudes mostly through the operator $Q_6$ in Eq.~(\ref{quarkeffH}). 
    \item We compute this imaginary part using $\left<B^\prime|{\cal Q}_6|B\right>$ bag model matrix elements followed by leading order $\chi$PT to obtain the corresponding S and P waves.
    \item  To estimate the uncertainty in this extraction of the imaginary part, we use the one-loop corrections in $\chi$PT. It is noteworthy that this estimate of the uncertainty covers the range of early model predictions that relied on simple hadronic models.
    \item The real part of the amplitudes are taken from experiment.
\end{itemize}
This procedure results in the weak phases for $\Delta I=1/2$ amplitudes,
\begin{align}
   (\xi^S-\xi^P)_{\Lambda\to p\pi^-} =&(-0.1\pm1) ~{A^2\lambda^5\eta}\\
   (\xi^S-\xi^P)_{\Xi^-\to \Lambda \pi^-} =&(1.5\pm1.1) ~{A^2\lambda^5\eta},
\end{align}
which in combination with the data in Table~\ref{ta1} and the strong phases quoted below that table, leads to the CP observables:
\begin{align}
    A_{CP~SM}^\Lambda\sim& (-3{\rm ~to~}3)\times 10^{-5},\\
    A_{CP~SM}^\Xi\sim &(0.5{\rm ~to~}6)\times 10^{-5},\\
    B_{CP~SM}^\Xi\sim &(-3.8{\rm ~to~}-0.3)\times 10^{-4}.
\end{align}

\subsection{Weak phases beyond the SM}

The estimate of these phases once again involves large uncertainty as it is the same calculation outlined for the SM case but with a new physics operator containing the dominant CP phase replacing $Q_6$. To arrive at numerical estimates we will make some simplifying assumptions:
\begin{itemize}
    \item The NP does not affect in a significant way the real part of the amplitudes.
    \item There is only one dominant operator that carries the new phase.
    \item The new CP violating phase occurs in a $|\Delta S|=1$ operator thus contributing to both $\epsilon^\prime$ and $\epsilon$ (in combination with a SM weak transition through long-distance contributions).
     \item We then use the new physics windows shown in Table~\ref{t0} to constrain the NP phases.
\end{itemize}

\begin{figure}[!htb]
\centerline{\includegraphics[width=1.0\linewidth]{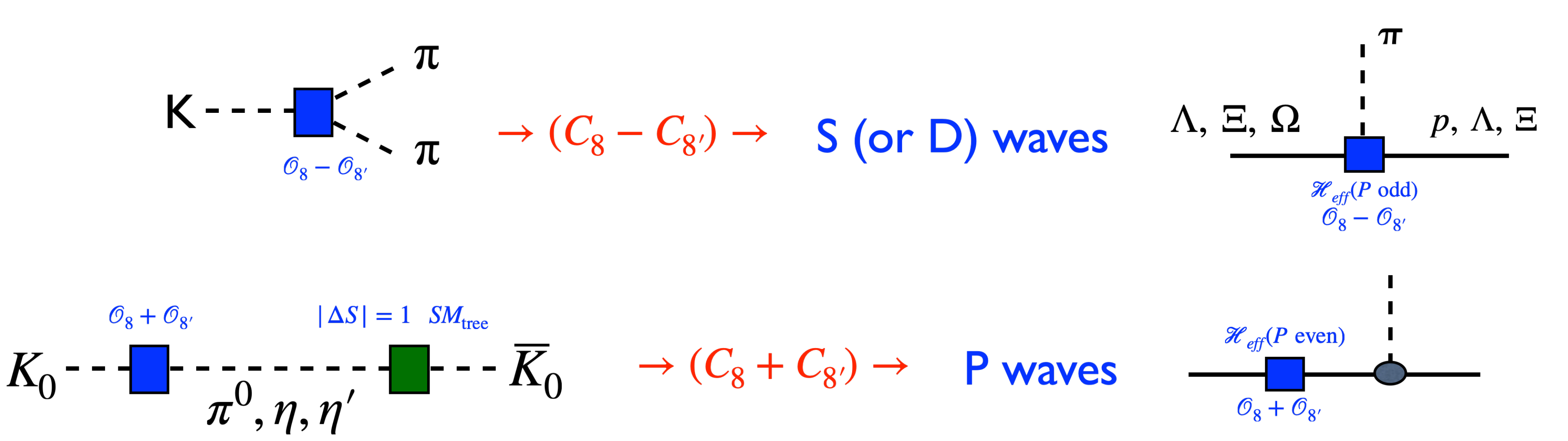}}
\caption{Illustration showing how the kaon observables constrain the BSM weak phases in hyperon non-leptonic decay.\label{f3}}
\end{figure}

To proceed in a model-independent manner, \cite{He:1995na} considered CP phases in all the $\Delta S=1$, dimension six, SMEFT operators from \cite{Buchmuller:1985jz}. The contributions of these operators to $\epsilon$, $\epsilon^\prime$ and $A_{CP}^\Lambda$ were then estimated using vacuum saturation for the matrix elements and it was found that two operators were the most promising candidates to enhance CP violation in hyperons while satisfying the kaon constraints:
\begin{align}
    &\bar d_Rs_L\bar u_R u_L +{\rm h.c.} \\
    &\overline d_{L(R)}\sigma^{\mu\nu}T^as_{R(L)}G^a_{\mu\nu} +{\rm h.c.}
\end{align}
The second one of these operators appears in many models, and was used in the early estimates of \cite{Donoghue:1985ww,Donoghue:1986hh} in the context of the “Weinberg model”. A more detailed study in the context of a supersymmetric model \cite{He:1999bv} later emphasized the complementarity of the hyperon and kaon modes. We will use this operator for the following estimate with the current notation 
\begin{align}
    {\cal L}_{NP}&\supset {\color{red}C_8}{\cal O}_8+{\color{red}C_{8^\prime}}
    {\cal O}_{8^\prime},\\
    {\cal O}_{8(8^\prime)}=&\frac{4G_F}{\sqrt{2}}V_{ts}V_{td}^*\frac{g_s}{16\pi^2}m_s\overline d_{L(R)}\sigma^{\mu\nu}T^as_{R(L)}G^a_{\mu\nu}
\end{align}
Fig.~\ref{f3} illustrates the complementarity between the three observables in constraining the new phases in $C_8$ and $C_{8^\prime}$. While $\epsilon^\prime$ constrains the phase in the parity odd combination of operators that contributes to the S-wave in octet hyperon non-leptonic decay, $\epsilon$ constrains the parity even combination that enters the P-waves of octet hyperon non-leptonic decay.\footnote{In a similar manner, kaon mixing constraints the real part of parity even $|\Delta S|=2$ transitions \cite{He:2023cqg}.} Note that this contribution to kaon-mixing and the parameter $\epsilon$ is through long-distance contributions as sketched in Fig.~\ref{f3}.

\begin{figure}[!htb]
\centerline{\includegraphics[width=1.0\linewidth]{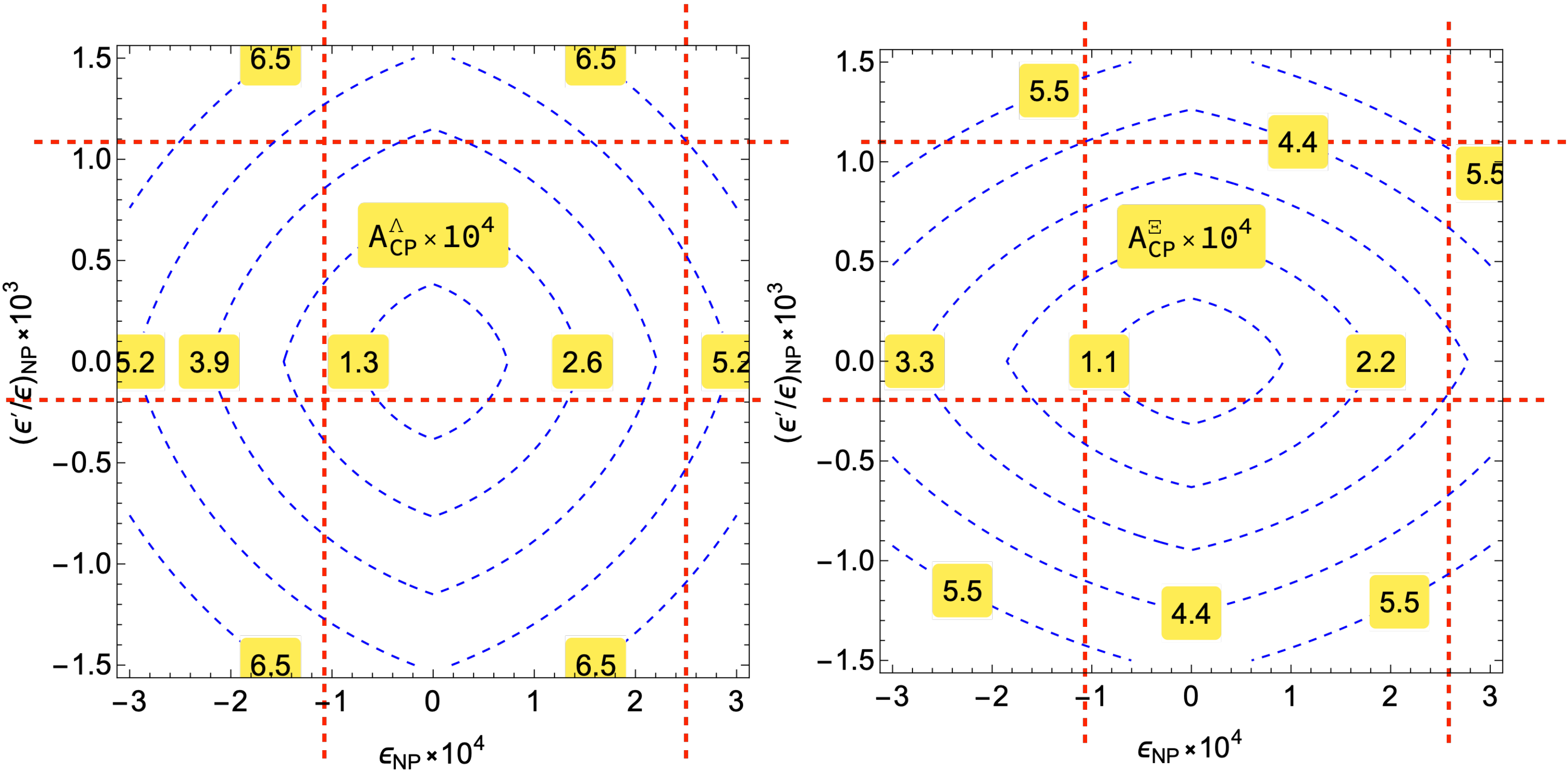}}
\caption{Contours of $A_{CP}^\Lambda$ and $A_{CP}^\Xi$ as functions of the amount of $\epsilon$ and $\epsilon^\prime$ that is attributable to new physics through the operators $O_{8,8^\prime}$. The red dotted lines mark the limits from Table~\ref{t0}.\label{f4}}
\end{figure}

The correlation between these observables is further illustrated in Fig.~\ref{f4}. The figure displays contours of $A_{CP}^\Lambda$ and $A_{CP}^\Xi$ as functions of the amount of $\epsilon$ and $\epsilon^\prime$ that is attributable to new physics through the operators $O_{8,8^\prime}$. The red dotted lines mark the limits from Table~\ref{t0}. Within these limits we see that  $A_{CP}^\Lambda$  could reach values of up to $\sim 6\times 10^{-4}$. Coincidentally, this was the same conclusion reached in \cite{He:1999bv} for certain SUSY scenarios even though the status of both $\epsilon$  and  $\epsilon^\prime$ have changed.

\begin{figure}[!htb]
\centerline{\includegraphics[width=1.0\linewidth]{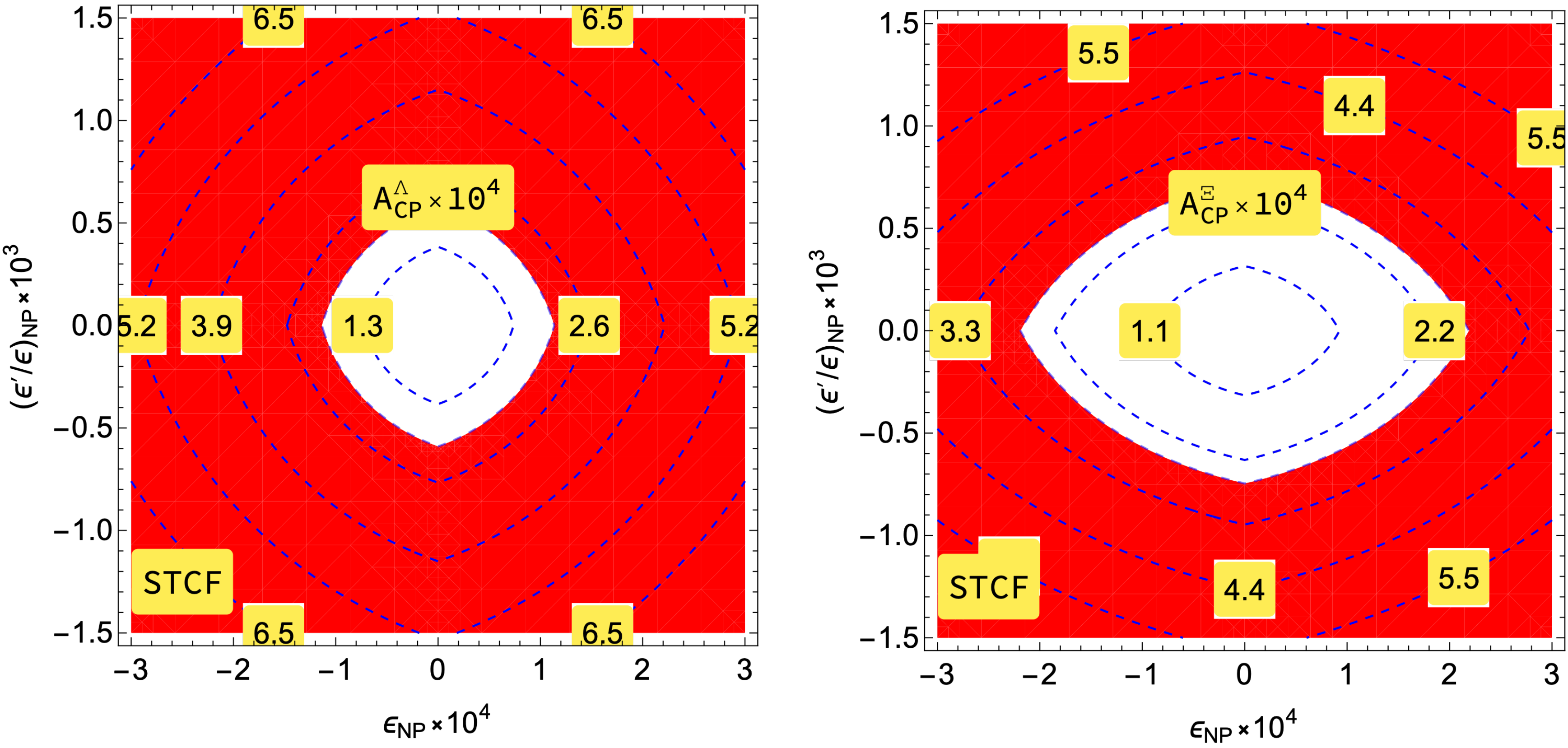}}
\caption{Region that could be excluded with the STCF projected sensitivity with $3.4\times 10^{12} ~J/\psi$ (in red) superimposed on Fig.~\ref{f4}.\label{f5}}
\end{figure}

In Fig.~\ref{f5} we superimpose in red on Fig.~\ref{f4} the region that could be excluded with the STCF projected sensitivity with $3.4\times 10^{12} ~J/\psi$ to $A_\Lambda~(A_\Xi)\sim 2 ~(2.6)\times 10^{-4}$. 

\begin{figure}[!htb]
\centerline{\includegraphics[width=0.53\linewidth]{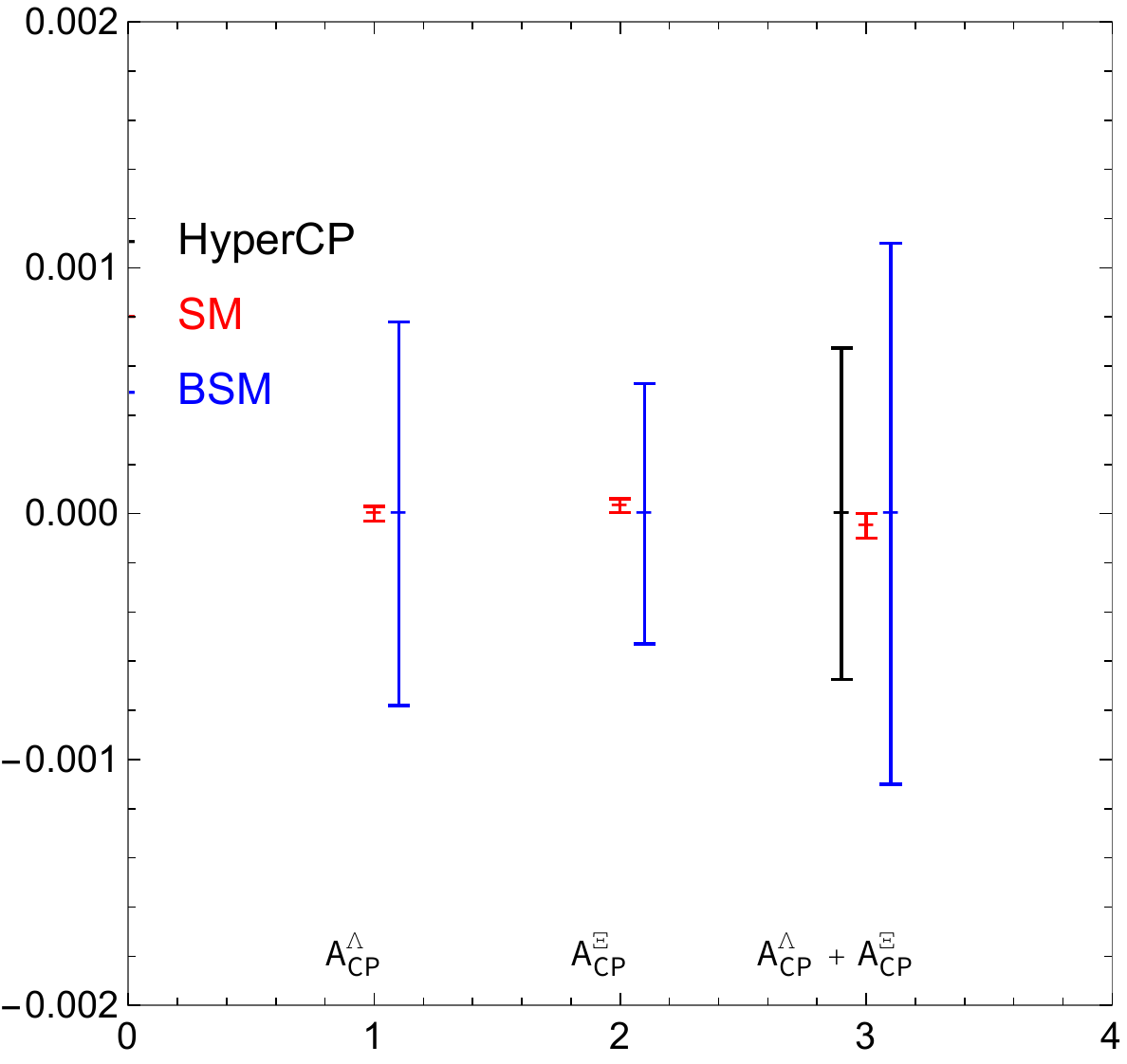}~~\includegraphics[width=0.45\linewidth]{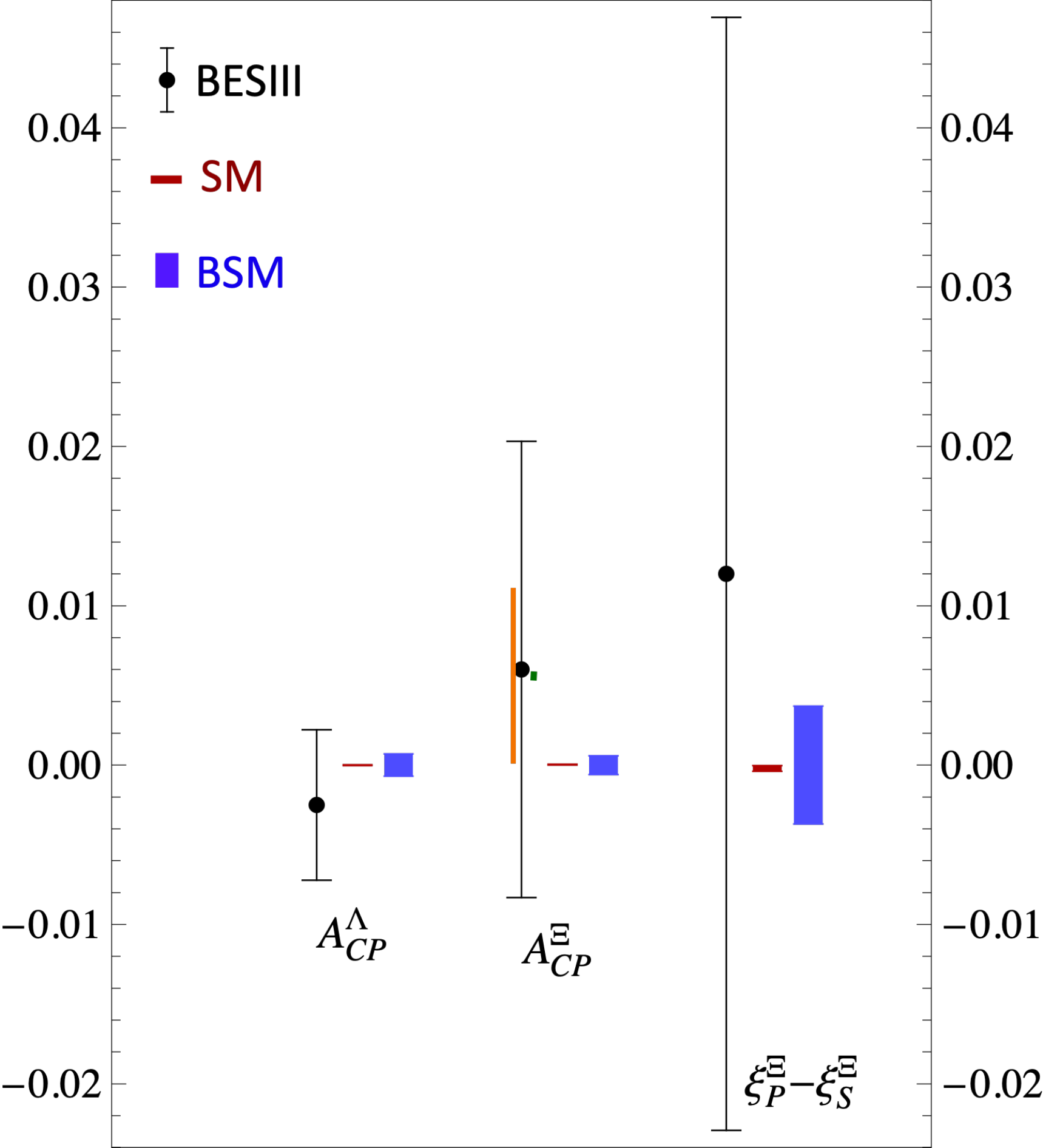}}
\caption{Comparison of SM and BSM through $O_{8,8^\prime}$ ranges of selected CP observables to the measurement by HyperCP (left panel) and the current BESIII limits (right panel). On the right panel, for $A^\Xi_{CP}$, we have added the projected statistical sensitivities expected from BESIII (orange line) and STCF (green dot) according to Table~\ref{ta3}.\label{f6}}
\end{figure}

A slightly different view was taken in \cite{Tandean:2003fr,Tandean:2004mv,Salone:2022lpt,He:2022bbs} where the weak phases for hyperon decay were written as
\begin{align}
    (\xi_P-\xi_S) \sim \left(C^\prime\left(\frac{\epsilon^\prime}{\epsilon}\right)_{BSM}+C~\epsilon_{BSM}\right).
\end{align}
Combining this with $\left|\frac{\epsilon^\prime}{\epsilon}\right|_{BSM}\lesssim1\times 10^{-3}$ and $\left|\epsilon\right|_{BSM}\lesssim 2\times 10^{-4}$, it is then found that \cite{Salone:2022lpt}
\begin{align}
    |A_{CP}^\Lambda|&\lesssim 7\times 10^{-4},\\
|A_{CP}^\Xi|&\lesssim 5.9\times 10^{-4},\\
|B_{CP}^\Xi|&\lesssim 3.7\times 10^{-3}
\end{align}
These numbers can be used to compare the estimates within the SM and beyond to the measurements by HyperCP (left panel) and BESIII (right panel) in Fig.~\ref{f6}. For $A_{CP}^\Xi$ in the right panel we show in black the current BESIII limit, and in orange (slightly to the left) and green (slightly to the right) projected the statistical sensitivity in future BESIII and STCF as per Table~\ref{ta3}.

\section{Conclusions}

Hyperon decays can play an important role in probing BSM physics in the $s\to d$ sector complementing kaon decays. To be competitive with existing $\epsilon$ and $\epsilon^\prime$ results, much higher sensitivity is needed and this may be achieved at a STCF.
Hyperon decay modes allowed in the SM receive large long-distance contributions that are difficult to estimate reliably. We have presented existing estimates for CP-violating observables that exhibit a large uncertainty. For further theoretical progress, the involvement of the lattice community is essential. 
Recent BESIII measurements have significantly increased our knowledge of CP-violating observables in hyperon decay and we look forward to their future improvements. 
A super tau-charm factory with $10^{12}-10^{13}~~J/\psi$ leading to $10^9-10^{10}$ reconstructed hyperon decays has the potential to test CP violation at levels near those estimated for the SM and to cover much of the BSM window.

\section*{Acknowledgments}

This work was supported in part by the Australian Government through the Australian Research Council. This talk was based on work done in collaboration with Xiao-Gang He and Jusak Tandean.

\bibliography{biblio}
\end{document}